\documentstyle[amssym,aas2pp4,epsfig,flushrt]{article}
\begin{document}
\newcommand{\Mo}{\mbox{$\rm M_\odot$}}
\newcommand{\Lo}{\mbox{$\rm L_\odot$}}
\newcommand{\mic}{\mbox{$\rm \mu m$}}
\newcommand{\ivol}{\mbox{$\rm cm^{-3}$}}
\newcommand{\isup}{\mbox{$\rm cm^{-2}$}}
\newcommand{\isec}{\mbox{s$^{-1}$}}
\newcommand{\Av}{\mbox{$A_{\rm V}$}}
\newcommand{\Ak}{\mbox{$A_{\rm K}$}}
\newcommand{\Ne}{\mbox{$N_{\rm e}$}}
\newcommand{\Np}{\mbox{$N_{\rm p}$}}
\newcommand{\Te}{\mbox{$T_{\rm e}$}}
\newcommand{\ten}[1]{\mbox{$10^{#1}$}}
\newcommand{\xten}[1]{\mbox{$\times 10^{#1}$}}
\newcommand{\wl}{\mbox{$\lambda$}}
\newcommand{\forb}[2]{\mbox{$[{\rm #1\, #2}]$}}
\newcommand{\Ha}{\mbox{H$\alpha$}}
\newcommand{\PA}{\mbox{Pa$\alpha$}}
\newcommand{\Hb}{\mbox{H$\beta$}}
\newcommand{\QH}{\mbox{$Q(\rm H)$}}

\lefthead{Capetti et al.}
\righthead{HST Infrared imaging polarimetry of Cen A}

\title{HST infrared imaging polarimetry of Centaurus A:
implications for the unified scheme and the existence of a mis-directed
BL Lac nucleus\footnote{Based on
observations with the NASA/ESA Hubble Space Telescope, obtained at the
Space Telescope Science Institute, which is operated by AURA, Inc.,
under NASA contract NAS 5-26555 and by STScI grant GO-3594.01-91A}}

\author{Alessandro Capetti}
\affil{Osservatorio Astronomico di Torino\\
       Strada Osservatorio 20, I-10025 Pino Torinese, ITALY} 

\author{Ethan J. Schreier}
\affil{Space Telescope Science Institute\\
       3700 San Martin Drive, Baltimore, MD 21218, USA}
       
\author{David Axon, Stuart Young, J. H. Hough, Stuart Clark}
\affil{Division of Physics and Astronomy, Dept. of Physical Sciences,\\ 
University of Hertfordshire, College Lane, Hatfield, Herts AL10 9AB, UK}
       
\author{Alessandro Marconi}
\affil{Osservatorio Astrofisico di Arcetri\\
       Largo E. Fermi 5, I-50125 Firenze, ITALY}

\author{Duccio Macchetto\altaffilmark{2}}
\affil{Space Telescope Science Institute\\
       3700 San Martin Drive, Baltimore, MD 21218, USA}

\and

\author{Chris Packham}
\affil{Isaac Newton Group, Sea Level Office\\
Apartado de Correos, 321, 38780 Santa Cruz de La Palma,Islas Canarias, SPAIN}

\altaffiltext{2}{Associated with Astrophysics Division, Space Science Dept.,
ESA}
 
\begin{abstract} 

We report results from HST/NICMOS 2\mic\ imaging polarimetry of the central
region of Centaurus A. 
In the vicinity of the nucleus we observe a complex polarization structure
which we explain by a combination of scattering of nuclear light 
and dichroic polarization associated with
the dust lane.  The scattered nuclear
radiation is found in an angular region which extends over $\gtrsim
70^{\circ}$ and thus it does not originate from a highly collimated
beam, but is associated with more omni-directional nuclear
illumination. 
These observations also show the presence of an
unresolved, highly polarized (P = 11.1\%) 
nuclear source whose polarization angle $\theta = 148.2^{\circ}$ 
is perpendicular to the jet axis. We set an upper limit of $0\farcs04
(\sim$0.8 pc) to its extent. 
The observed nuclear polarization is naturally accounted for 
if we are observing scattered light from an otherwise obscured nucleus
provided that both the scattering region and the occulting torus 
are extremely compact, with an outer radius of less than $\sim 1$ pc.
Alternatively, we might be directly seeing the infrared counterpart of the 
radio core, similar to those found in other low luminosity radio-galaxies 
observed with HST.
We discuss these results in the framework of the FR~I / BL~Lac unifying model.

\end{abstract} 


\section{Introduction}

Centaurus A (= NGC5128) is the nearest active galaxy, and both the AGN and
the jet have been the subject of extensive studies in all wavelength bands
from radio through gamma rays.  Radio observations provide strong upper
limits on the size of the central source - at 0.5 $\pm$ 0.1 milliarcsec
(0.01pc), it is the smallest known extragalactic radio source (Kellerman,
Zensus \& Cohen 1997). Because of the heavy obscuration due to the dust
lane, it is very difficult to study the AGN itself at high spatial
resolution in the optical. Ground-based IR observations have provided
evidence for a strong, highly polarized source at the nucleus (cf.
\cite{bailey:86}, \cite{packham:96}) and Bailey et al. suggested that Cen A
contained a low luminosity, mis-directed BL Lac source.
Further evidence for this hypothesis came from the conclusions 
of Morganti et al. (1991),
that the blue filaments in the jet are photo-ionized by a relativistically
beamed continuum source. {\sl Conversely, Antonucci and Barvainis (1990)
attributed the observed polarization to scattering
of nuclear light.} While similarly arguing
for scattering as the source of the polarization, Packham et al. (1996)
suggested that Cen A could still be a BL Lac type object. 

High resolution polarimetry of this nearest AGN is important not only to
the understanding of AGN in general, but to verifying the unified model
and more specifically, the suggestion that Fanaroff-Riley type I radio
galaxies provide the parent population for BL Lac objects (cf. Urry \& Padovani
1995). Previous Hubble Space Telescope (WFPC-1) R band imaging polarimetry
of the inner region of NGC5128 (\cite{schreier:96}) 
failed to reveal any details of the polarization structure at the nucleus 
because of its extreme reddening but identified a small 
region of polarization with a scattering knot southwest of the
nucleus.
Subsequent Near Infrared Camera and Multi-Object
Spectrometer (NICMOS) observations revealed the existence of 
a nuclear point source which could also be identified 
in the optical images of the Wide Field Planetary Camera (WFPC2) 
allowing its flux to be determined from V through K
(Marconi et al. 2000).   
NICMOS narrow band imaging in Pa $\alpha$ and FeII 
lead to the discovery of 
an elongated structure which was interpreted as a 20 pc radius ionized 
gas circumnuclear disk
(Schreier et al. 1998). 

We report here HST 
(NICMOS) polarimetric observations at 2\mic\, which allow us to map the
polarization structure in the vicinity of the nucleus at much higher
resolution (0\farcs3) than previously possible.  
HST's improved spatial resolution
over the ground, and NICMOS's sensitivity and polarimetric capability at
2\mic\ allow us to isolate the polarized nuclear component, and to
study the effects of the AGN on its nearby environment. We summarize the
observations and data reduction in Section 2 and the analysis and observed
results in Section 3.  The interpretation of the off-nuclear polarization
structure is presented in Section 4. The
origin of the polarized nuclear emission, and the
implications for the AGN unified scheme, are
discussed in Sections 5 and 6. 
Summary and conclusions are given in Section 7.
Throughout, we assume a distance to
Centaurus A of 3.5 Mpc (\cite{hui:93}), whence 1\arcsec$\simeq$17pc. 

\section{Observations and Data Reduction}

The nuclear region of NGC 5128 was observed on May 6, 1998 using NICMOS
Camera 2 with the long wavelength polarizers. 
The bandpass of the polarizers covers the
spectral range 1.9 - 2.1 \mic\ and their principal axes
of transmission are
nominally oriented at PA 0, 120 and 240 degrees. The spatial resolution is
0\farcs075/pixel and the field of view of the camera is 19\farcs4
$\times$ 19\farcs4 (256 $\times$ 256 
pixels.)  All observations were carried out with a MULTIACCUM sequence
(\cite{mackenty:97}), i.e. the detector is read out non-destructively
several times during each integration to facilitate removal of cosmic rays
and saturated pixels.  The data were re-calibrated using the pipeline
software CALNICA v3.0 
(\cite{bushouse:97}) and the best reference files in the Hubble Data
Archive to produce flux calibrated images.  Bad pixels were removed
interpolating from values of neighboring pixels.  

The observing strategy used varying integration times from 16s to
960s for each polarizer, to allow for potential saturation,
caused by the bright IR nucleus.
Repeating the observation sets with all polarizers in each of 4 orbits,
allowed the removal of time dependent effects. The highest count rate observed
is 330 c/s per pixel; even for the longest exposure times, the total counts
only reach $\sim 12000$ in each of the 25 individual read-outs, well below
the saturation limit of $2\times$ $10^5$ and within the linearity regime of 
the camera. Thus, all the observations are
within the linear regime of the camera. This is confirmed by comparing the
short and long exposure images, which reveal no significant differences in
observed count rates.  We present results from the analysis of the set of
12 images obtained with the longest (960s) exposure times, one for each of
the three polarizers, taken in four subsequent orbits. 

A drift in the bias level is known to be present in the NICMOS images and
this results in spatially dependent residuals in the calibrated images (the
``pedestal'' problem, \cite{pedestal}). We estimated the effects of the 
pedestal on our observations by comparing images taken with 
each polarizer in different
orbits. The differences between the images are always smaller than
$\sim$ 0.5 \%.
We thus estimate that the presence of the pedestal could
translate into a spurious polarization on an otherwise unpolarized source
of less than 0.6 \%. 

A bright star is clearly visible in the field of view, $\sim 8 \arcsec$
southwest of the nucleus, as in all previous HST images of Cen A.  We used
the position of this star to check the alignment of the individual images.
Shifts between images taken with the different polarizers are negligible,
$\lesssim$ 0.03 pixels.  There is a small drift between images taken in
different orbits, amounting to a total of 0.12 pixels from the first to the
fourth orbit.  The point-spread functions in the three polarizers do not
show significant differences.

After registration, images were co-added and combined to produce the final
polarization maps. Preflight thermal vacuum tests 
showed that the transmission of the polarizers is not identical
and that they are offset from their nominal position angles
(Hines, Schmidt \& Lytle 1997).
Polarization parameters were then estimated using the
algorithm developed by Sparks and Axon (1999) which deals with this
non ideal instrumental configuration and which allowed us to derive the 
Stokes parameters and, finally, the polarization images.

The vacuum tests also showed that any instrumental NICMOS 
polarization is $\lesssim 1 \%$.  On-orbit calibration observations of the
polarized star CHA-DC-F7 produce values that agree with 2 \mic\ ground
based observations within 0.2\% at the 1\% polarization level.  

\section{Results}

\begin{figure}
\centering
\psfig{figure=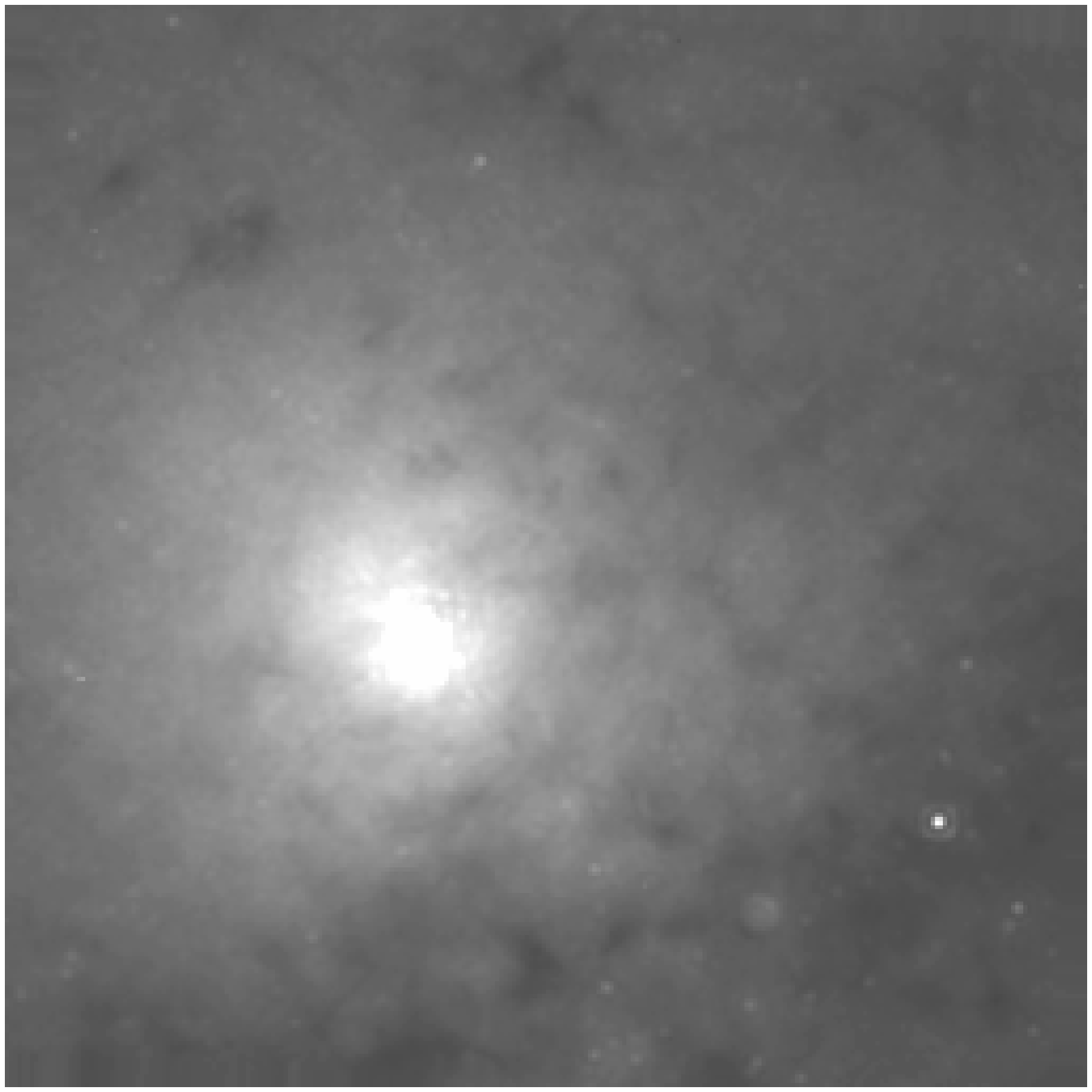,width=0.9\linewidth}
\vskip 0.5cm
\caption{\label{fig:i} Gray scale image
of the nuclear region of Cen A in the wavelength range 1.9 -- 2.1 \mic. 
The field of view is 19\arcsec $\times$ 19\arcsec. For all images, North is up and east is to the left.} 
\end{figure}

Fig. \ref{fig:i} shows the total intensity map over the entire NICMOS
field of view. Not surprisingly, given the similarity between the long
wavelength polarizers spectral band and the medium band F222M ("K-band")
filter, this image displays the same basic features as the NICMOS image
presented by Schreier et al. (1998). A central unresolved
source is superimposed on a smooth galactic structure, punctuated by
regions of high absorption that are particularly evident toward the South
where the thickest regions of the dust lane are located.

\subsection{The off-nuclear polarization}

\begin{figure*}
\centering
\psfig{figure=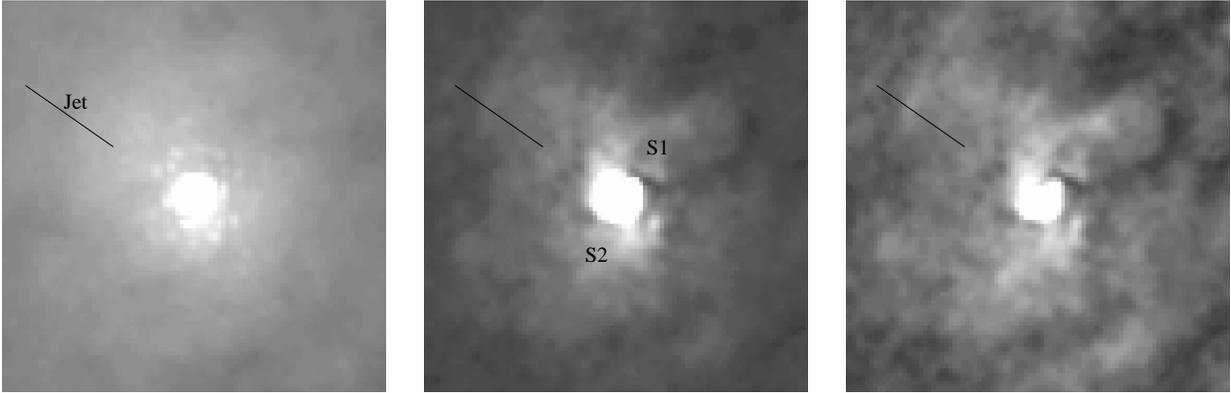,width=0.3\linewidth,angle=-90}
\vskip 0.5cm
\caption{\label{fig:ipm} Gray scale total intensity image of the innermost
7\arcsec $\times$ 7\arcsec of Centaurus A. 
Dots surrounding the nucleus are artifacts of the NICMOS PSF. 
From left to right the images are total intensity, polarized intensity
and percentage of polarization.
The long dash marks the radio-jet orientation.} 
\end{figure*}

In Fig. \ref{fig:ipm} we present, from left to right, maps of the total
intensity, polarized flux and percentage polarization, respectively, in the
central 7\arcsec $\times$ 7\arcsec.  The polarized flux clearly shows a
more complex structure than the total intensity image.  Two elongated
features extend out from opposite sides of the nucleus along PA $\sim
25^{\circ}$ to a distance of about 1\arcsec, 
where the polarization reaches 4\%.  
These features,
together with two fainter extensions oriented approximately perpendicularly
to the previous ones, suggest an S-shaped structure centered on the
nucleus; this structure is more apparent in the 
percentage polarization map. The rest of the polarized emission is rather
diffuse, with polarization ranging between 2 and 3 \%.

\begin{figure}
\centering
\epsfig{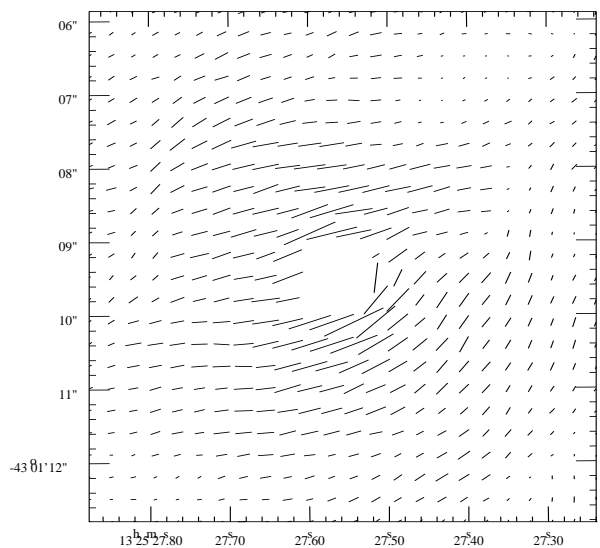}
\vskip 0.5cm
\caption{\label{fig:vec} Polarization vectors
 of the innermost 7\arcsec $\times$ 7\arcsec of Centaurus A.
Vectors are plotted every 0\farcs3\ and the length is proportional 
to the polarized flux.
The central source has been masked out
to show more clearly the off nuclear polarization pattern.} 
\end{figure}

Fig. \ref{fig:vec} shows the polarization field for the same central
7\arcsec\ $\times$ 7\arcsec\, re-binned at 4 $\times$ 4 pixels (0\farcs3
$\times$ 0\farcs3); the lengths of the vectors are proportional to the
polarized flux and the orientation of the vectors indicates the polarization position angle.  Outside the central source, the overall polarization
pattern is dominated by a uniform structure oriented approximately along
the dust lane.  The integrated polarization of the central 2\farcs25, after
subtracting the nuclear
component, is oriented at PA $\sim$ 106$^\circ$, with $P$ = 2.6\%.  This is
consistent with the ground based measurements and with the HST/WFPC1 value
for the polarization angle, $\sim$ 110$^\circ$.  However, there are clearly
large deviations from a uniform field which suggest a centro-symmetric
structure. 

\begin{figure}
\centering
\psfig{figure=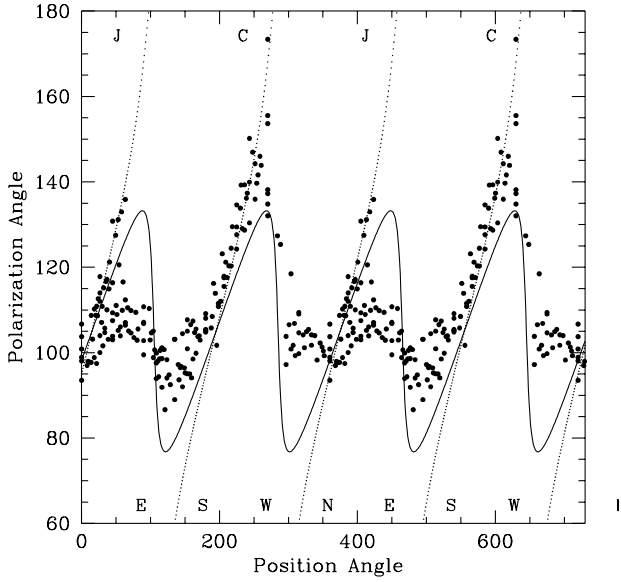,width=1.0\linewidth}
\vskip 0.5cm
\caption{\label{fig:model} Polar diagram of the polarization position angle 
with the origin centered on the nuclear IR source. Two periods are plotted
for clarity. Deviations from a uniform pattern (oriented at PA $\sim 105$)
are seen and are centered at the position angle of the jet and counter jet.
Lines represent polarization models
resulting from the superposition of a centrosymmetric and a uniform pattern 
for two values of their intensity ratio $R$ = 2 (dotted line) and $R$ = 0.5
(solid line).} 
\end{figure}

To quantify this effect, we plot in Fig. \ref{fig:model} the polarization
angle at each pixel versus the position angle (plotted twice for clarity).
As usual, position angle 0 corresponds to North and the angle increases
towards East. In addition to a uniform polarization field oriented at
PA$\sim$105$^\circ$, large deviations with a quasi-periodic behavior
suggest that an additional component or components are present. We suggest
that the dominant effect can be understood as a nuclear-symmetric pattern
superimposed on the uniform field. We discuss the origins of the extended
polarization components in Section 4 below.

\subsection{\label{sec:nucpol} Nuclear polarization and photometry }

We performed aperture photometry on each of the 12 long exposure images
separately, to estimate the polarization parameters of the nucleus and to
derive direct estimates of the statistical errors. The total flux density
of the nucleus is 21.2 $\pm$ 0.2 mJy (1.6 \xten{-15} erg cm$^{-2}$ s$^{-1}$
\AA$^{-1}$). Its percentage polarization is 11.1$\pm$0.2\% and the
polarization position angle is 148.2$\pm$1.0$^\circ$. Systematic errors are
estimated at $\lesssim$ 5\% for the total intensity, due to the uncertainty
in the NICMOS calibration, while
the effects of instrumental polarization and pedestal are negligible on
such a highly polarized source.

The size of the nuclear infrared source is estimated by fitting the radial
profiles of the nucleus and the bright star SW of the nucleus as a
comparison. The FWHM are 2.30 $\pm$ 0.02 and 2.32 $\pm$ 0.01 pixels for the
nucleus and the star respectively (errors derived from the spread of the
12 independent images). We conclude that the central IR source in Cen A is
unresolved at HST resolution. An upper limit to its extent can be
determined assuming that the PSF and the nuclear angular dimension add in
quadrature. The three-sigma upper limit for the nuclear FWHM is then 0.6
pixels, or 0.8 pc. The polarized source is also unresolved, with a formal
FWHM = 2.34 $\pm$ 0.06 pixels; the corresponding three-sigma upper limit on
its size is 1.2 pixels, or 1.4 pc.

A direct comparison with previous (ground-based) IR imaging polarimetry of
Cen A cannot be performed since the NICMOS polarizers bandpass differs
significantly from standard infrared filters. The mismatch is particularly
important due to the steepness of the Cen A nuclear infrared 
spectrum (Marconi et al. 2000).
We note, however, that within a 2\farcs25 synthetic aperture, our values
for the polarization parameters fall between those measured by Packham et
al. (1996) in the H and K bands, bracketing the wavelength range of our 
observations (see Tab. 1). 

\section{ The extended components: scattering and the dust lane }

The polarization map presented above suggests a superposition of two
components - a constant field and a centro symmetric component. 
The field resulting from this superposition 
depends on the relative intensity $R$ of the two polarizing
mechanisms. If the uniform field dominates, the reflection pattern would
produce periodic oscillations around the mean value, whose amplitude
increases with $R$. In the limit of pure scattering (no uniform field), we
would expect straight lines representing an offset of 90 degrees with
respect to the polar angle.  In Fig. \ref{fig:model} we illustrate two
models, for $R = 2$ (dotted line) and for $R= 1/2$ (solid line),
respectively. Note that the intensity ratio $R$ of the two components is
the only free parameter; the average value of the uniform polarization
field can be fixed from ground based observations. 
Clearly, the largest deviations
from uniform polarization, which occur around PA$\sim50$ and $\sim250$ are
well fitted by models in which the scattering field dominates, i.e. for
large values of $R$. 

\begin{figure}
\centering
\psfig{figure=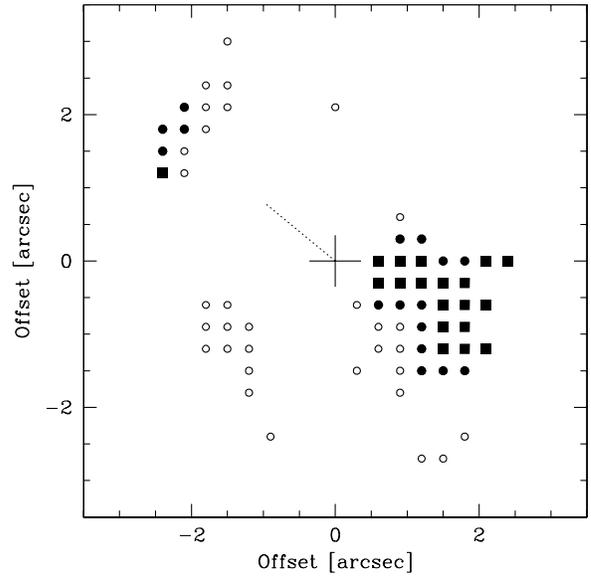,width=1.0\linewidth}
\vskip 0.5cm
\caption{\label{fig:dev} Location of the largest deviations of the 
polarization field from the uniform pattern. Empty circles are deviations
between 10 and 20 degrees, filled circles between 20 and 30, squares
above 30 degrees.
The dashed line marks the radio-jet orientation.}  
\end{figure}

In Fig. \ref{fig:dev} we plot the deviations from constant polarization
angle as a function of position with respect to the nucleus. The deviations
range up to 70$^\circ$, and the largest occur from PA 35$^\circ$ to
60$^\circ$ and from PA 190$^\circ$ to 280$^\circ$. Much smaller amplitude
deviations of $\sim$ 15$^\circ$ degrees are present at PA 130$^\circ$, and
no significant deviations are seen in the fourth quadrant (PA
$\sim$300$^\circ$). This shows that significant scattering occurs in the
nuclear region of Cen A, in a broad bi--cone
centered on the jet axis (PA $\sim$55$^\circ$).

The two elongated features seen in the polarization images, which form the
body of the S-shaped structure we pointed out in Section 3.1
(labeled S1 and S2 in Fig. 2b) are
not aligned along the position angle of the jet (at the VLBI scale,
PA=51$^\circ$, Tingay et al. 1998). 
On the other hand, they are co-spatial with the small,
ionized gas disk discovered in Pa $\alpha$ by Schreier et al. (1998)
and the polarization vectors are perpendicular to the radius vector to the 
nucleus.  
The most likely explanation of these structures is therefore
that they are the result of a concentration of scattering material 
associated with the
disk of ionized gas rather than being directly associated with
a collimated nuclear beam.

The faint outer extensions of the S-shape structure show a rather different
behavior than the inner parts: 1) they are not associated with line
emission, and 2) their polarization angle is aligned with the large-scale
dust lane.  Both results can be accounted for if these features lie outside
the region illuminated by the nucleus, with their polarization therefore
being accounted for by dichroic transmission rather than scattering.

\section{\label{sec:nucorigin} Origin of the polarized nuclear emission}

In this section, we examine three possible explanations for the highly
polarized source observed at the nucleus of Centaurus A - 1) dichroic
transmission through dust, 2) scattering, and 3) synchrotron radiation - as
other authors have previously done (cf. Packham et al. 1996). In the next
section, we discuss the implications of our results. 

\subsection{\label{sec:dichroic} Dust dichroic transmission}

If we assume the observed polarization arises from dichroic transmission
through aligned dust grains, we can calculate the implied extinction. We
empirically compare our data to observations of the center of our own
galaxy, where a polarization of 6.4\% at K is seen (Bailey, Hough \& Axon
1984), and the extinction is \Av $\sim$ 30 mag.  We conclude that \Av
$\sim$ 50 mag (i.e. \Ak $\sim$ 6.6 mag) 
is required to produce the observed polarization of 11.1\% at
the nucleus of Cen A. {\sl The intrinsic 2\mic\ flux of Cen A would then be
of $\sim 9$ Jy. As a comparison, the mid-infrared nuclear flux of Cen A
at, e.g., 15 \mic\, is 1.2 Jy (Mirabel et al. 1999). 
Such a large flux increase toward shorter wavelengths 
is not expected regardless on the origin of the nuclear emission.
In fact, broad band spectra of radio quiet quasars 
(Barvainis 1990), which we take as 
representative of the case in which 
the nuclear emission is thermally dominated, as well as 
optically thin synchrotron emission, show the opposite behaviour.
The discrepancy with the predicted intrinsic Cen A spectrum is 
sufficiently large that cannot simply be ascribed to, e.g., a higher
efficiency of dichroic polarization and leads us to rule out this 
explanation.}

\subsection{\label{sec:scattering} Scattered nuclear light }

If the nuclear polarization originates from scattering of light
from an otherwise obscured nucleus, our
observations put a strong constraint on the size of the 
scattering region and the obscuring torus, both of which must have an 
outer radius smaller than 0.8 pc.
To test this possibility we have used the
scattering model presented in Young (2000) to determine if the implied
compact region can scatter enough flux to account for the observed
unresolved source. The model, itself an expansion of the model of Young
et al. (1995), allows for a conical {\sl electron} 
scattering region within a torus-like
geometry of a flared disk.  The extinction through the torus from any
point within the scattering region is calculated in-situ.

The outer radius of the scattering cone was taken to be 0.8 pc, the largest
size compatible with our unresolved source.
The inner and outer radii of the torus were chosen to be 0.03 pc and
1 pc, respectively. The model included a vertical scale-height for
the dust density away from the torus equatorial plane, taken to be 1 pc. 

Typically for active galaxies with obscured broad line regions 
about 1 per cent of the incident flux is scattered into the line of sight
(Miller and Goodrich 1990; Young et al. 1996). 
With an inner scattering radius of 0.03 pc, and an extinction
through the torus of A$_V$ = 47 magnitudes along the direct line of sight, it
was possible to reproduce the observed degree of polarization while
maintaining an optically thin scattering region with a number density of
scatterers of $3 \, 10^{5}$ cm$^{-3}$ at the inner scattering radius.
Such a density 
is comparable to that implied from the scattering modelling of some 
IRAS galaxies (Young et al. 1996).

While the inner radius of obscuring tori in AGN is set
by the dust sublimation radius, the appropriate value for its outer radius
is matter of debate but it has been usually taken between 10 and 100 pc
(Young 2000 and references therein), 
much larger than the limit set by our observations.
However, modelling of the Spectral Energy Distribution (SED) of Cen A
also suggests a quite compact torus, with an outer radius of $\sim$ 3.6 pc 
(Alexander et al. 1999), when compared to other sources 
(e.g. Efstathiou, Hough \& Young 1995). Furthermore the radio observations
by Jones et al. (1996) require the presence of dense material
confined to within 1 pc from the nucleus. Taken together these results 
imply that any obscuring torus in Cen A must be
very compact and surrounded by a high density scattering cloud.

\subsection{\label{sec:synchro} Synchrotron emission}

Analysis of a complete sample of HST images of Fanaroff-Riley type I
(FR~I) radio galaxies shows a strong linear correlation between optical and radio
core emission (Chiaberge, Capetti \& Celotti 1999). The average
radio-to-optical spectral index of the sample, defined by the best-fit
correlation, is 0.75. The correlation extends over four orders of magnitude
in luminosity, with a dispersion of only 0.4 dex.
This lead Chiaberge et al. to argue for a 
common, non-thermal synchrotron origin for both the radio and optical
emission. 

The very high column density estimated from soft X-ray
photoelectric absorption of $9.42\times 10^{22}$ cm$^{-2}$ (Rothschild et
al. 1999) is often thought to be too large to allow a direct view 
of the nucleus of Cen A at optical wavelengths. In fact, 
it corresponds to a visual extinction of $A_V\sim
47$ mag, assuming a standard gas-to-dust ratio $A_V=5\times 10^{-22} N_H$
{\sl which would completely obscure the nuclear source in the optical
and, as discussed above, require an untenably high intrinsic infrared flux}.
However, Granato, Danese \& Franceschini (1997) argue that most of the
X-ray absorption in optically obscured AGNs is produced within the dust
sublimation radius, where the gas is essentially dust free.  A conversion
of column density to extinction based on the galactic gas-to-dust ratio may
thus lead to an overestimate of $A_{\rm V}$. A value of only one third
standard would reduce the extinction to $A_V \sim15$ mag, making the optical
nucleus directly observable even in the optical band.  
We note this value agrees with that
derived from the 10\mic\ silicate absorption feature
and with the observations of the nuclear spectral energy
distribution by Marconi et al. (2000).

In this context, the radio core
flux for Cen A of 9.1 Jy at 15 GHz (Clarke, Burns \& Norman 1992)
predicts a flux of 10 mJy at 2\mic \,
which, seen through a reddening of \Av$\sim$15 mag, 
is only 0.7 dex smaller than our measurement of 21 mJy.
{\sl Similarly, the optical V flux of 6.3 \xten{-20} erg cm$^{-2}$ s$^{-1}$
\AA$^{-1}$ (Marconi et al. 2000), 
when corrected for such an amount of extinction falls
again only 0.7 dex below the FR~I radio/optical nuclear correlation 
\footnote{As discussed in detail by Marconi et al.  
the central wavelength of the F555W image, 5440 \AA, 
is significantly offset from its effective value, 5990 \AA, due to the 
steepness of Cen A nuclear spectrum, wavelength 
at which fluxes and extinction must by then evaluated.}.}

Potentially, a more substantial problem for the synchrotron interpretation 
of the continuum is that while the nucleus of Cen A is heavily polarized in 
the IR, it is unpolarized at 800 and 1100 \mic\ 
(with a 3 $\sigma$ upper limit of 0.7 \% at 1100 \mic\ Packham et al. 1996)
{\sl and it shows little polarization in the radio band (0.46 \%,
Burns, Feigelson and Schreier, 1983). 
However, a similar behaviour is seen in the two FR~I nuclei 
for which nuclear polarization measurements exist 
from HST observations. 
In M~87 the nucleus is polarized at 2.9 \% in the UV (Capetti et al. 1995)
while it shows a polarization level of only 0.95 \% 
in the radio (Eric Perlman, private communication); similarly, analysis 
of archival UV images of 3C~264 revealed a polarization of 23 \%, 
while the radio polarization is only 3.5 \% (Lara et al. 1997). 
All polarization angles are,
within the errors, consistent with being perpendicular to the 
respective radio jet axis, and this is the case also for Cen A
whose radio polarization is oriented at PA = 149 (Rusk and Seaquist 1985).
The observed IR polarization properties
of the nucleus of Cen A nicely fit in this scheme which suggests a common
synchrotron origin for the polarization in the radio and 
in the IR/optical/UV bands.
Certainly intriguing is the location and origin of the medium responsible for
the depolarization observed at radio wavelengths.}

\section{Implications for the FR~I / BL Lac unified model}

In this section we explore the implication of our results for Cen A 
as a representative member of the FR~I population in the framework of 
the unifying model for low luminosity radio-sources.
The role of obscuration in this scheme is still a matter of debate.
Although
circumnuclear tori are crucial for the unification of broad and narrow line
Seyfert galaxies
and powerful radio sources, there is as yet no evidence in favour
of their existance in FR~I. Indeed they are not required implicitely
for the FR~I / BL Lac unification (see Urry and Padovani 1995).
The results of Chiaberge et al. 
suggest that tori are either geometrically thin (height over 
radius $\simeq 0.1$) or present only in a minority of FR~I.

It is clear from our previous discussion that our IR polarization 
observations of Cen A can be well explained in a model in which the nucleus is 
surrounded by an obscuring torus
and the infrared nuclear source (as well as its optical counter part)
is purely due to scattering. 
By allowing the scattering region to be partially obscured, 
the extinction observed between
the V-band and the near-IR can also be reproduced.  

Alternatively, as discussed in the previous section, 
we might have a direct view of the synchrotron emission from the
nucleus of Cen A.
Packham et al. (1996) pointed out that the nucleus of Cen A has rather
different properties from BL Lac objects. In particular, they noted: a) lower
polarization at mm wavelengths than typical for BL Lacs; and b) a
polarization angle which is approximately perpendicular to the jet axis,
while in BL Lacs it is usually parallel to it. 
In addition our observations do not show 
a highly collimated polarized region cospatial with the jet axis
which again argues against the presence of a misdirected
BL Lac nucleus in Cen A. 

However convincing these arguments appear 
when taken together with our discussion about scattering, 
they do not strictly rule out a synchrotron origin for the
nuclear polarization, since there is growing evidence that the emission 
which dominates in BL Lacs
does not originate from the same region of that seen in FR~I. First,
the polarization of radio cores in all radio galaxies is much lower than in
BL Lacs (Saikia 1999). 
Second, the intensity ratio between BL Lacs and FR~I
nuclei is typically $10^3 - 10^4$, indicative of Lorentz factors $\Gamma
\sim 5 $ again smaller that typical values for BL Lacs, i.e.
$\Gamma \sim 10 - 15$ (Ghisellini et al. 1993, Chiaberge et al.
2000). In fact Chiaberge et al. concluded that the core emission in FR~I comes
from a less anisotropic component than in BL Lacs, due to the
presence of a velocity structure within the jet, which allows slower
components to dominate the faster ones at large viewing angles.
A similar conclusion has been derived by Laing et al. (1999)
discussing the jet asymmetry of FR~I. 
If we adopt this view then the nuclear polarization of Cen A can
still be due to synchrotron from an unresolved slow jet halo,
surrounding a high $\Gamma$ jet.

A key constraint provided by our observations is that one has to 
explain not only the nuclear polarization but also the nature 
of the off nuclear scattered light.
While this is readily accounted for if there is a shadowing torus,
in the synchrotron picture the polarized light is reflecting 
the intrinsic anisotropy of the nuclear radiation field.
Our images show that the angular region covered by scattered light has an 
opening angle of $\sim 70-90$$^\circ$ and therefore requires only 
modest anisotropy in the nuclear emission.
The opening angle (FWHM) of the illuminating beam is
related to the jet bulk Lorentz factor as $\theta \sim 2 \Gamma^{-1}$.
If this is due to Doppler beaming it requires 
$\Gamma \sim 2$ in agreement with the estimates of Lorentz factors
for the jet halo derived by Chiaberge et al. (2000).

Because the source of illumination is intrinsically polarized
in principle one should expects deviations from circular symmetry
in the polarization pattern. 
The detailed modeling of this effect shows that the source polarization
must be comparable with the maximum polarization caused by scatterers 
(60 \%) 
in order to have a significant impact on the observed pattern.
Unfortunately in Cen A the foreground dichroic sheet prevents any such 
effect to be seen.

However, we also have to explain the absence of a collimated
scattered counter part to the fast synchrotron core of the radio jet.
The only obvious way to do it is to make
this region essentially devoid 
of scatterers by clearing them with the passage of the arcsec scale jet.

\section{Summary and conclusions}

The large scale polarized component seen in the nuclear regions of Cen A
is consistent with previous ground-based and HST WF/PC-I
observations and is explained by dichroic transmission through the
foreground dust lane. Significant quasi-periodic deviations in the
polarization angle within a few arcseconds of the nucleus 
are caused by scattering of nuclear radiation. 
The strongest effects are seen in the quadrants that
contain the jet symmetry axis, but there is no detailed correlation with
the jet itself. 

The nucleus is found to be unresolved, with a 3-sigma upper limit on
the FWHM of 0.8 pc. It is highly polarized, at 11.1$\pm$0.2\%, 
with a position angle of 148.2$\pm$1.0$^\circ$, perpendicular to the jet axis.

The nuclear polarization cannot be explained by
dichroic transmission through dust, as this would require an absorption of 
\Av$\sim$50 mag. {\sl The intrinsic 2 \mic\ flux would then exceed by a factor 
$\sim$ 10 the mid infrared measurements and this is not expected
regardless on the origin (thermal or non thermal) 
of the nuclear emission.}

Our modeling shows that it is possible to explain the nuclear 
polarization as optically thin scattering from a $\sim 1$ pc cone.
This requires the existence of an extremely compact torus.
However, this is in agreement with the modeling of the SED of the Cen A 
nucleus and also with the radio observations.
This model also naturally explains the large scale biconical scattering
morphology.

A synchrotron origin for the nuclear radiation requires  
that it originates from a low $\Gamma \sim 2$ 
unresolved jet halo surrounding the faster jet core so that the angular size
of extended scattering can be matched.
This kind of jet morphology is expected on the basis
of recent results on the statistical properties of FR~I radio-galaxies.
In order to explain the absence of a bright scattered BL Lac beam,
i.e. a high $\Gamma$ jet core, one has to assume that the jet 
clears away the scattering material along its trajectory. 

It would be very important at this stage 
to establish whether circumnuclear tori are common in FR~I
or whether their observed properties are dominated by their jet
emission. Unfortunately,
previous measurements of the nuclear polarization in FR~I have always 
been hampered by the dominant dilution of starlight 
as their nuclear sources only account for a few per cent 
of the total flux in typical ground based apertures 
(see e.g. Impey, Lawrence \& Tapia 1991).
A high resolution HST study of the sort present here for the FR~I class 
as a whole is needed to establish the polarization 
properties of their IR and optical cores and their pattern of illumination.
 
\acknowledgements
We would like to thank the referee, R. Antonucci, for his useful comments 
and suggestions.

\newpage

\newpage

\begin{table}{\label{tab:photo}}

\caption{Comparison of HST and ground based polarimetry}

\begin{tabular}{c l c  c c c} 

 & & & & & \\
\hline
 & & & & & \\
 & Band &   $\rm{I_P}[\%]$ & $\theta$ & I [mJy] & P [mJy]\\
 & & & & & \\
\hline
 & & & & &  \\
 & H   &  3.45 &  129.7 &  73 &   2.5 \\
 & HST &  3.83 &  135.6 &  69 &   2.6 \\
 & K$_n$   &  5.43 &  142.2 &  88 &   4.8 \\
 & & & & & \\

\hline 
\end{tabular}
\end{table}

\end{document}